\newcommand{\st}{ \mbox{SrTiO$_3$} }
\newcommand{\bt}{ \mbox{BaTiO$_3$} }
\newcommand{\kn}{ \mbox{KNbO$_3$} }
\newcommand{\kt}{ \mbox{KTaO$_3$} }
\newcommand{\beq}{\begin{equation}}
\newcommand{\eeq}{\end{equation}}
\newcommand{\noi}{\noindent}
\newcommand{\nup}{ \mbox{$n\uparrow$}  }
\newcommand{\ndn}{ \mbox{$n\downarrow$}  }
\newcommand{\exc}{ \mbox{$\epsilon_{xc}^{unif}$} }

\documentstyle[preprint,aps,12pt]{revtex}
\begin{document}
\title{Applications of the generalized gradient approximation to
 ferroelectric perovskites}

\author{S. Tinte$^*$, M.G. Stachiotti$^*$, C.O. Rodriguez$^{**}$, 
         D.L. Novikov$^\dagger$ and N.E. Christensen$^\ddagger$}
\address{ $^{*}$ Instituto de F\'{\i}sica Rosario, Universidad Nacional de
              Rosario, \\ 27 de Febrero 210 Bis, 2000 Rosario, Argentina \\
         $^{**}$ IFLYSIB, Grupo de F\'\i sica del S\'olido, C.C.565, 
              1900 La Plata, Argentina \\
         $^\dagger$ Science and Technology Center for Superconductivity,
          Department of Physics and Astronomy, Northwestern University, 
                    Evanston, IL, 60208,USA \\
         $^\ddagger$ Institute of Physics and Astronomy, Aarhus University,
                    DK-8000 Aarhus C, Denmark.}

\maketitle

\begin{abstract}

The Perdew-Burke-Ernzerhof  generalized gradient 
approximation  to the density functional theory 
is tested with respect to sensitivity to the choice of the value of the
parameter $\kappa$, which is
 associated to the degree of localization of the exchange-correlation
hole. A study  of  structural and dynamical properties of four selected
ferroelectric perovskites is presented.    
The originally proposed value of $\kappa$=0.804
%(best suited for atoms and molecules)
works well for some solids, whereas                
for the ABO$_3$ perovskites it must be 
decreased in order to predict equilibrium lattice 
parameters in good agreement with experiments. The effects on the 
structural instabilities and zone center phonon
modes are examined.
The need of varying $\kappa$ from one system to another
reflects the fact that the localization of the exchange-correlation
hole is system dependent, and the sensitivity of the structural
properties to its actual value illustrates the necessity of
finding a universal function for $\kappa$.

\end{abstract}

\section*{Introduction}

Due to their wide variety of physical properties, the ABO$_3$
perovskites constitute a particularly interesting family for studies 
of structural phase transitions. These depend strongly
on the composition of the compounds, as for instance on the 
chemical nature of the A and B ions in pure samples, as well as on the 
relative composition in the case of solid solutions. Furthermore,
the temperature driven structural phase transitions of a given pure system 
are strongly affected by the crystal volume and thus by application of         
external pressure.    

The electronic, structural and dynamical properties of ferroelectric 
perovskites have been the subject of intense theoretical work, but only 
recently has first principles density functional theory been applied with 
success to analyze them at a microscopic level.
The local (spin-)density approximation, L(S)DA, has provided a useful framework
to elucidate important aspects of the underlying physics of these oxides, 
providing important quantitative information about 
electronic charge distributions, character of electronic bondings, crystal
structure, phonons, structural instabilities, etc. 
For example, frozen phonon and linear response approaches have been 
applied to the study of the lattice dynamics of many perovskites, such as
\kn \cite{pos94,pos942,sin92,yu95}, \bt~\cite{coh90,gho97},
\kt \cite{pos94,sin96}, \st \cite{las97}, etc ;
providing considerable insight into the nature of the soft-mode
total energy surface.

The main drawback of the LDA when  applied to perovskites is the 
systematic underestimation of the equilibrium volume, and several properties 
are poorly reproduced at this volume. 
In fact, in prototypical materials, like \kn\ and \bt, ferroelectricity is
strongly suppressed at the LDA equilibrium volume. 
Therefore, L(S)DA studies often use a simple {\it ad hoc} correction                
 consisting in using       
the experimental rather than the calculated equilibrium volumes. 
This is just one example of shortcomings that have
 motivated a number of attempts to go beyond the LSDA in the 
search of a better framework. The application of the Perdew and Wang (PW91) 
version \cite{pw91} of the Generalized Gradient Approximation (GGA) 
functional to \kn\ and \bt\ 
showed that the equilibrium volumes of the two compounds are 
overcorrected; although in \kn\ GGA yielded  an equilibrium volume
that is much closer to the experimental volume than the LSDA result, 
in \bt\ the GGA error was almost as large as found within the LDA. \cite{sin95}   
More recently, the Weighted Density Approximation (WDA) has been applied 
to \kn\ yielding an equilibrium volume in close  agreement with experiment.
\cite{sin97}

In the Kohn-Sham density functional theory only the exchange-correlation
energy E$_{XC}=$E$_X+$E$_C$ which is a functional of the electron spin 
densities must be approximated, and for slowly 
varying densities, $n$, it can be expressed  as the volume integrals of 
$n$ times \exc in the LSDA case and 
$f$(\nup,\ndn,$\nabla$\nup,$\nabla$\ndn) for the GGA case.
For practical calculations the exchange-correlation energy density      
of a uniform electron gas, \exc(\nup,\ndn), and $f$ must be   
parametrized. The form of \exc is now well established but 
which is the best choice of $f$ is still under debate.

In the recent work of Perdew, Burke and Ernzerhof (PBE) \cite{pbe96}
a simplified form of the GGA-$f$ was presented which improves over the
previous PW91
in an accurate description of the linear response of the uniform gas,
behaving correctly under uniform scaling and  giving a smoother potential.
The new PBE retains the correct features of LSDA and combines them with 
the most energetically important features of gradient corrected nonlocality.
Studies of atomization energies for small molecules \cite{pbe96} gave 
essentially the same results as PW91.

However, a remainig uncertainty relates to the value of the parameter   
$\kappa$, in the  enhancement factor F$_X(s)$ 
which is directly associated to the degree of localization of the 
exchange-correlation hole. (Here the variable $s$ is a measure of the
relative density gradient, { $s$ = $|$$\nabla$$n$$|$/2$k{_F}$$n$ } 
$k_{F}$ giving the Fermi wavenumber of an electron gas of density $n$).
In their original work PBE proposed
F$_X(s)= 1+ \kappa -\kappa / (1+\mu s^2/\kappa)$
which satisfied the inequality 
F$_X \leq 1.804$ 
with $\kappa =0.804$ and with the value of $\mu \cong 0.21951$.
Zhang and Yang found \cite{zhang98} that the results for several
atoms and molecules were improved by increasing $\kappa$ beyond the
originally proposed value, 0.804. But, this does not hold for all
type of bonds\cite{pbe98}, and  it
may well happen that applications to solids appear to be improved
when smaller values of $\kappa$ are used.
The parameter $\kappa$ might  be a weak function of the reduced Laplacian,
$\kappa$=g($\nabla$${}^2$n/(2k${}_F$)${}^2n$),
which would also satisfy the condition 'D' 
in the PBE paper of uniform density scaling,  
i.e. that E$_X$ scales like $\lambda$.

A study of LSDA and PBE-GGA ($\kappa$=0.804) has recently been made for
$s-p$ materials \cite{martin97}, 3$d$ and 4$d$ transition metals 
\cite{kok98} and another, similar to the present, also including 
magnetic effects, has been made for the 3$d$, 4$d$ and 5$d$ transition 
metal series \cite{peltzer98}.

In the present work we take four different compounds as representative 
examples of the ferroelectric perovskites in their cubic phases.
The PBE proposal of the GGA for the exchange correlation energy 
within DFT is tested against its intrinsic uncertainty in choosing
the coefficient $\kappa$ and also compared to the LSDA results.
For this we have selected three structure related properties which relate to 
 the physics of the ferroelectric 
instabilities and phase transitions:
(1) the energy-volume curves from which lattice parameters and bulk moduli 
are derived,  
(2) $\Gamma_{15}$ phonon modes for the cubic structure, 
and
(3) the tendency to suffer a ferroelectric transition when the 
atoms are displaced according to the soft-mode displacement pattern.

The selected perovskites are  \kn, \bt, \st and \kt,     
since they exemplify the rich variety of structure related 
and dynamical effects of the perovkite materials.
These oxides have been extensively studied because of their possible
technological applications, some of which are related to their ferroelectric 
properties. But a basic understanding of the mechanisms of their various 
transformations of phase and structure is also of great interest.

The two compounds, \kn and \bt, exhibit the same sequence of phase
transitions upon cooling from the high-temperature cubic paraelectric
phase to slightly distorted ferroelectric structures with tetragonal,
orthorhombic, and rhombohedral symmetry. Their
transition temperatures have an 
approximately constant ratio : T$(\kn )/$T$(\bt )=1.6.$     
Thus the dynamics of these two
materials should be similar but with different energy scales.
SrTiO$_3$ has the simple cubic structure at high temperatures
going through an antiferrodistortive transition at 104~K to a 
tetragonal phase in which the oxygen octahedra are rotated in opposite
senses in neighboring unit cells. KTaO$_3$ remains cubic perovskite down to 
low temperatures. These two last materials present a softening of a transverse
optic phonon which correspond to the ferroelectric mode, but its frequency 
remains finite at all temperatures. This fact induced these two compounds 
to be termed {\em incipient ferroelectrics}.

\section*{Computational details}

The calculations presented in this work were performed within the 
LDA and the PBE-GGA to
density functional theory, using the full-potential
Linear Augmented Plane Wave (LAPW) 
method. In this method no shape approximation on either the
potential or the electronic charge density is made. We use the WIEN97
implementation of the method \cite{wien97} which allows the inclusion
of local orbitals (LO) in the basis,
improving upon linearization and making possible a consistent treatment
of semicore and valence states in one energy window
hence ensuring proper orthogonality \cite{LO}.
Ceperley and Alder \cite{cepal} L(S)DA exchange-correlation contribution,
as parametrized by Perdew and Zunger\cite{perzun}, was used.

The muffin-tin sphere radii ($R_i$)= 2.0, 2.0, 2.0 ,1.95, 1.90, 1.90
and 1.50 a.u. were used for K,   Ba,   Sr, Ti,   Nb,    Ta, and  O,
respectively. 
The value of the parameter $RK_{max}$, which controls the size of the basis
sets in these calculations, was chosen to be  8 for all  systems studied. 
This gives well converged
basis sets consisting of approximately
1200  LAPW functions plus local
orbitals for the different systems.
In order to obtain sufficient accuracy with inclusion of the GGA functionals
a relatively high plane wave cut-off energy (256~Ry) was necessary 
in order to obtain a converged interstitial representation
of the potential.
This choice of parameters was justified by performing
calculations for other $R_i$ values
and by increasing  $RK_{max}$. Although an increase of
$RK_{max}$ from 8 to 9 had no significant effects
on most of physical
properties that we studied,
convergence of the cohesive energy to a precision of 1~mRy per formula
unit required $RK_{max}$=10 with the choice of $R_i$ given above.

We introduced  LO to include the following orbitals in the basis set:
K-3$s$ and 3$p$,
Ba-5$s$, $5p$ and 4$d$, 
Sr-4$s$ and 4$p$,
Ti-3$s$ and 3$p$, 
Nb-4$s$ and 4$p$,
Ta-5$s$, 5$p$ and 4$f$ 
and O-2$s$.

Integrations in reciprocal space were performed using the special points
method. We used 6$\times$6$\times$6 meshes which represent 250 $k$-points
in the first Brillouin zone.
This corresponds to 28 special k-points in the irreducible wedge
for the rhombohedral structure, 18 for the tetragonal and 10 in the cubic.                                
Convergence tests indicate that only small changes result from 
increasing to a denser k-mesh.

Of the two allowed symmetries for $\Gamma $ point
phonon modes in the cubic perovskite structure we only studied the 
triply degenerate $\Gamma_{15}$ modes. There are three of these, 
not counting the zero frequency acoustic mode.
These are infrared active and include the "ferroelectric soft" mode.
We determined the phonon frequencies and polarizations for this particular 
symmetry by calculating atomic forces for several small displacements
( $\sim 0.01$~{\AA} ) consistent with the symmetry and small enough to
be in the linear regime. From the force as a function of displacement
the dynamical matrix was constructed and diagonalized.

\section*{Results}

Table I lists the lattice parameters and bulk moduli as derived from
energy-volume curves obtained within the LDA as well as the PBE-GGA (with
$\kappa=0.804$ as originally proposed by PBE \cite{pbe96}),        
 together with
the experimental values. 
The equilibrium volumes obtained from the LDA calculations 
are 4.3\%, 5.0 \% , 2.3 \% and 2.0 \%
smaller than experiment for \kn, \bt, \st\ and \kt,  respectively.
This agrees with the generally observed tendency to overbinding in this
approximation. 
The LDA bulk moduli are overestimated when evaluated at these too small 
volumes. The values calculated at the experimental volumes (see the values 
in brackets) agree better with the experiments.

As is the general trend in the application of GGA functionals to solids,
it expands bonds, an effect that sometimes corrects and
sometimes overcorrects the LSDA prediction.
All GGA calculations with $\kappa$=0.804 overestimate the 
equilibrium volumes, namely by 1.6 \% , 1.6 \% , 4.2 \% and 4.0 \%
for \kn, \bt, \st\ and \kt, respectively.
We agree with previous calculations by Singh \cite{sin95} for \kn, but our 
results for \bt\ are different. Indeed, we obtained for both oxides that the 
GGA yields to an equilibrium volume that is much closer to the experimental
volume than the LSDA result. However, for the two incipient ferroelectrics
the GGA errors are twice as large as the LSDA ones.
We have not found any GGA calculations for  
\st\ and \kt\ to compare with.

In Fig. 1 the values of V/V$_o$ (V$_o$ is the  corresponding experimental 
value) are given for the four perovskites when the value of $\kappa$ in the 
PBE-GGA functional is decreased from 0.804 to 0.3. 
As can be seen both \bt\ and \kn\ would give perfect determination of the 
lattice parameter (V/V$_o$=1) for $\kappa \approx 0.6$. 
In the case of \st\ and \kt\ the value should be further reduced to 
$\approx 0.4$.
The need of varying $\kappa$ from one system to another
reflects the fact that the localization of the exchange-correlation
hole is system dependent. 
Nevertheless, it is important to investigate the sensitivity of 
the vibrational properties and energetics
which are involved in the ferroelectric instabilities 
by the application of the PBE-GGA functional and the reduction of the 
$\kappa$ value. 
This constitutes a hard test on the quality of any ab-initio scheme, since the
curvature of the total energy surface over the manifold of varios atomic
displacements is much more sensitive to the details of the calculations than
merely the position of the total energy minimun.

To clarify the effect of the GGA functional on the phonon energies, we have
performed frozen phonon calculations for the four selected perovskites at 
their experimental lattice constants, and examined the effects of choosing
 different exchange-correlation approximations,  
 i.e. LDA and PBE-GGA with
different $\kappa$ values.
The calculated frequencies of the $\Gamma_{15}$ modes and
the experimental values are shown in Table II.
Two  conclusions can be drawn from these calculations. 
Firstly,  the GGA functional hardens the phonon frequencies 
(as compared to the LDA frequencies) of the four perovskites.  
This hardening produces a slight reduction of the errors, 
since the LSDA provides phonon frequencies which are understimated 
by 10-20$\%$ compared with experiments.
The second conclusion concerns 
 the parameter $\kappa$. It is evident from 
Table II that the effect introduced on the GGA phonon frequencies by the 
modification of $\kappa$ is negligible.

This picture also holds for the low-frequency $\Gamma_{15}$ mode, 
i.e. the ferroelectric mode. In the cases of \kn\ and \kt, the GGA produces 
a small change of its frequency when it is compared with the LSDA result. 
However, a remarkable hardening is observed in the titanates 
where the GGA increases the imaginary frequency 
in \bt\  by $\approx$ 40 cm$^{-1}$ and stabilises the 
imaginary frequency mode in \st. In the latter material,    
the potential well where atoms move is very flat so the small differences 
in energetics between LSDA and PBE-GGA might give different 
predictions which would modify the underlying physics.

As a by-product of the frozen phonon calculation, we show in Table III the 
calculated eigenvectors of \kn\ to exemplify the effect produced on the 
phonon polarizations by the application of the GGA. 
Note that the eigenvectors are practically unchanged. The same occurs 
for the other three perovskites.

Finally, to test the sensitivity of the energetics involved in the 
ferroelectric instabilities when the different exchange-correlation 
functionals are used, 
we performed total energy calculations as a function
of the ferroelectric soft mode displacement pattern. 
In Figure 2, we show the energy as a function of the amplitude of (111) 
ferroelectric mode displacements for \kn\ and \bt. The calculations were
performed at the experimental lattice constants. In the case of \kn, 
the LSDA and GGA (with $\kappa$=0.804 and 0.6) yield a clear ferroelectric
instability with similar energetics and displacements, and with an energy 
lowering of $\approx$ 1.8 mRy/cell. 
However, for \bt\ the LSDA yields an energy 
lowering of 1.75 mRy/cell while the GGA yields 1.25 mRy/cell, which 
is consistent with the above mentioned hardening of the soft-mode frequency 
produced by the GGA.
The energetics in both materials are very little dependent on the $\kappa$ 
value used in the GGA functional. 
Although the transition temperatures are not associated directly with the
energy barriers shown in Figure 2, it is interesting to note that 
while for the LSDA the energy barrier ratio of the two compounds is  
E$(\kn )/$E$(\bt ) \approx 1$, for the GGA      
E$(\kn )/$E$(\bt ) \approx 1.5$ which is much closer to the      
transition temperatures ratio  T$(\kn )/$T$(\bt )=1.6.$

\section*{Conclusions}

Results of LSDA and PBE-GGA calculations have been reported for  
structural and dynamical properties relevant for the physics
of the ferroelectric instabilities of perovskites.
\kn , \bt ,\st\ and \kt were taken as representative example 
materials. 
In the case of PBE-GGA calculations were made with different values of
the coefficient $\kappa$ which is related to the localization
of the exchange-correlation hole.
The usual underestimation of LSDA for the equilibrium 
volume is obtained  (an average of approximately 3$\%$
for the four materials) in complete accord
with all other theoretical studies.
The PBE-GGA, on the other hand,  overestimates the equilibrium volumes 
when $\kappa$=0.804 (the originally proposed value) is used.
For $\kappa$ values of 0.6 in the cases of \kn\ and \bt\ and       
0.4 in \st\ and \kt\ a correct equilibrium volume (V/V$_o$=1.0)
can be obtained.
The need of varying $\kappa$ from one system to another
reflects the fact that the localisation of the exchange-correlation
hole is system dependent. 
However, the vibrational properties and energetics involved in the 
ferroelectric instabilities do not depend on the use of different $\kappa$
values. This fact would permit an {\it ad hoc} correction of the theoretical
equilibrium volume by selecting the adequate value of $\kappa$ for each
particular system.
However, this procedure would not lead to a fully ab-initio, free-paremeters, 
calculation scheme, indicating the necessity of finding a universal 
function for $\kappa$=g($\nabla$${}^2$n/(2k${}_F$)${}^2n$).

\acknowledgments
%\vspace{0.3cm}
%\noi { Acknowledgments} \\
This work was supported by
Consejo Nacional de Investigaciones Cient\'{\i}ficas y T\'ecnicas
de la Rep\'ublica Argentina,
the Danish Natural Science Research Council
(Grant No. 9600998) and the Commission of the European Communities,
Contract No. CI1*-CT92-0086.
M.S. also thanks Consejo de Investigaciones de la Universidad Nacional de 
Rosario for support.

%\newpage

\figure{Figure 1: Equilibrium volumes relative to the experimental ones of
\kn, \bt, \st  and \kt\ as a function of the PBE-GGA intrinsic
parameter $\kappa$. }

\figure{Figure 2: Total energy as a function of ferroelectric displacement 
along the (111) direction for \kn\ and \bt, calculated at the
experimental lattice constant for different exchange-correlation energies:
LSDA and PBE-GGA ($\kappa =$ 0.804 and 0.6).}

\newpage

\begin{table} 
\caption{Lattice parameter (in a.u.) and Bulk modulus (in GPa) obtained
for LSDA and PBE-GGA($\kappa =0.804$) and compared with experimental 
values. In parantheses: bulk modulus at the experimental equilibrium
volume.} 
\vspace*{0.3cm}         
\label{table1}
\begin{tabular}{cccccc} 
           & &   & LSDA & PBE-GGA ($\kappa =0.804$)    & Experiment \\ \hline
\\
KNbO$_3$   & & a & 7.48   &  7.63  & 7.59$^a$   \\
          & & B & 206(155)&  171(186)      & 138$^b$  \\
                                               \\
BaTiO$_3$  & & a & 7.44   & 7.61   &  7.57$^a$   \\
           & & B & 195(155)&  160(173)     &         \\
                                               \\
SrTiO$_3$  & & a & 7.30   &  7.46  &  7.358$^a$    \\
           & & B & 204(176)&  167(194)      & 179$^c$  \\
                                               \\
KTaO$_3$  & & a & 7.475   &  7.625  &  7.526$^a$   \\
           & & B & 222(192)&  183(213)      & 220$^d$  \\
                                               \\
\\
\end{tabular}
\end{table}
\noi $^a$ corresponds to the experimental equilibrium volume extrapolated
to T=0. \\
$^b$ calculated from cubic elastic constants, Ref. \cite{wie74}. \\
$^c$ Ref. \cite{mit69}. \\
$^d$ Ref. \cite{sam73}.

\newpage

\begin{table}
\caption{ Frequencies of the $\Gamma_{15}$ modes obtained for LSDA and PBE-GGA 
with two differents values of $\kappa$. $\kappa_{equil.}$ is the value of 
$\kappa$ for which the calculated equilibrium volume  agrees with  
experiment (see Fig. 1.)} 
\vspace*{0.3cm}         
\label{table2}
\begin{tabular}{ccccc}
Compound&  \multicolumn{4}{c}{Frequency (cm$^{-1}$)}         \\
\hline      
 & LDA & GGA ($\kappa$=0.804) & GGA ($\kappa$=$\kappa_{equil.}$)& Expt. \\ 
\hline \\
KNbO$_3$       & 211i  & 197i & 195i  & soft  \\
                & 166  & 182 & 179 & 198$^a$   \\
                & 466  & 478  & 478 & 521$^a$   \\  \hline
\\
BaTiO$_3$       & 239i  & 203i & 205i  & soft  \\
                & 163  & 168 & 167 & 182$^b$   \\
                & 454  & 463  & 461 & 482$^b$   \\  \hline
\\                
SrTiO$_3$       & 57i  & 64 & 52  & soft  \\
                & 157  & 166 & 166 & 175$^c$   \\
                & 532  & 551  & 545 & 545$^c$  \\  \hline
\\                
KTaO$_3$        & 99  & 107 & 109 & soft  \\
                & 172  & 189 & 181 & 197$^d$   \\
                & 529  & 550  & 545 & 546$^d$   \\  
\\
\end{tabular}
\end{table}
\noi $^a$ Infrared reflectivity measurements at 710 K, Ref \cite{fon84}. \\
$^b$ Infrared reflectivity measurements at 420 K, Ref \cite{lus80}. \\
$^c$ Infrared reflectivity measurements at room temperature,
Ref. \cite{ser80}. \\
$^d$ Hyper Raman scattering at room temperature, Ref. \cite{vog84}.\\

\newpage

\begin{table}
\caption{Frequencies and eigenvectors of the $\Gamma_{15}$ modes
in \kn\ using LSDA and PBE-GGA ($\kappa$=0.804 and 0.6). } 
\vspace*{0.3cm}         
\label{table3}
\begin{tabular}{cccccc} 
& Freq. (cm$^{-1})$ &  \multicolumn{4}{c}{Eigenvectors}                    \\
\hline      
     &  &  K         & Nb   & O1    & O2 and O3  \\ \hline
\\
KNbO$_3$       & 210i  & 0.015 &-0.58  & 0.6 & 0.39                  \\
LSDA, a=7.59au. & 166  &-0.88 & 0.36  & 0.146 & 0.178                    \\
                & 466  & 0.027 & -0.11 & -0.73 &0.48                    \\
\\
KNbO$_3$       & 196i  &0.042 &-0.59  &  0.57 & 0.40                   \\
GGA, a=7.59au. & 182  &-0.88  &0.35  &0.178  &0.18                   \\
$\kappa=0.804$  & 478  &0.008 &-0.08  &-0.75 &0.47                   \\
\\
KNbO$_3$       & 195i  &0.035 &-0.59  &  0.567 & 0.40                   \\
GGA, a=7.59au. & 179  &-0.88  &0.35  &0.17  &0.18                   \\
$\kappa=0.6$  & 478  &0.012 &-0.083  &-0.75 &0.46                   \\
\\
\end{tabular}
\end{table}

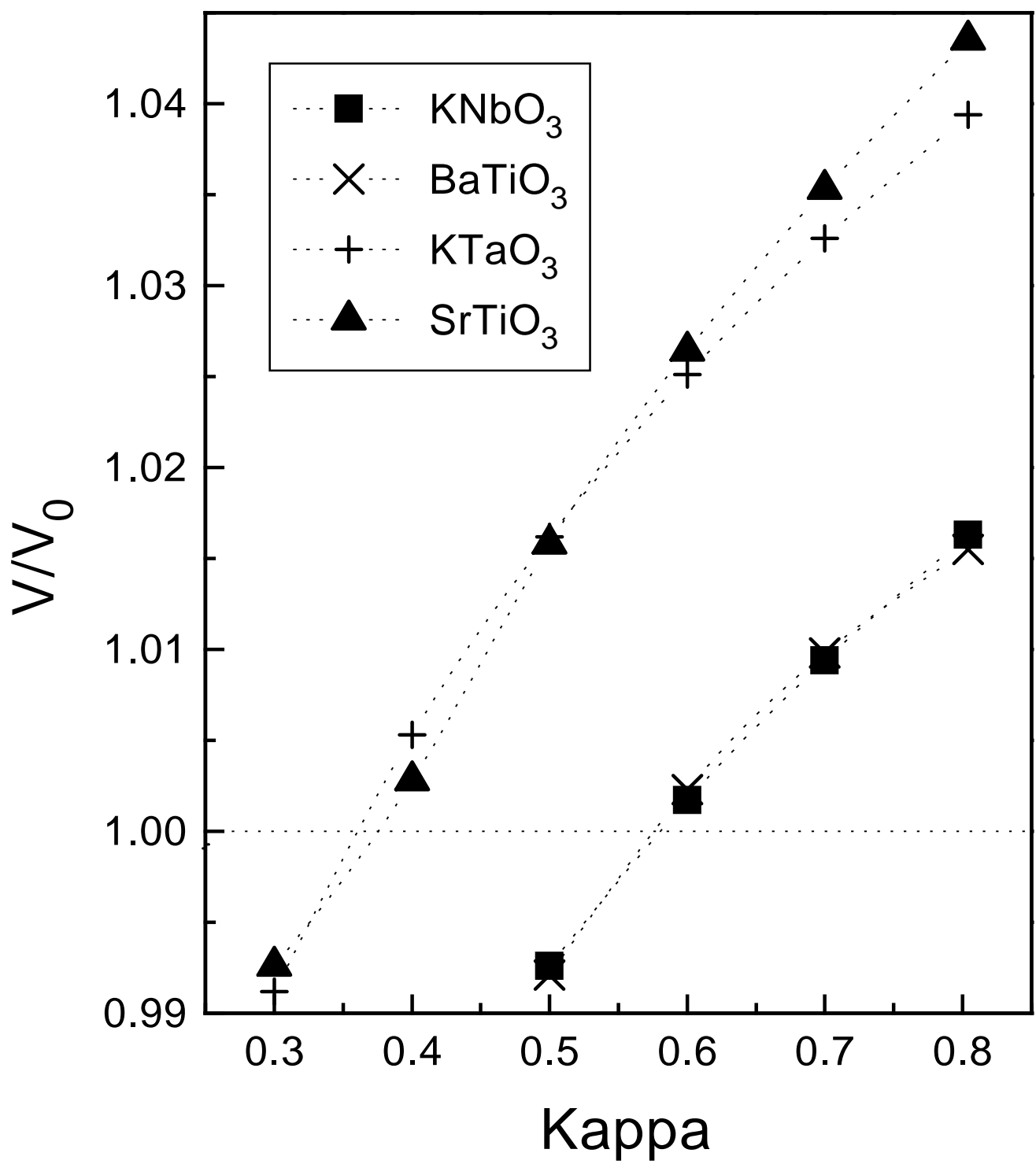
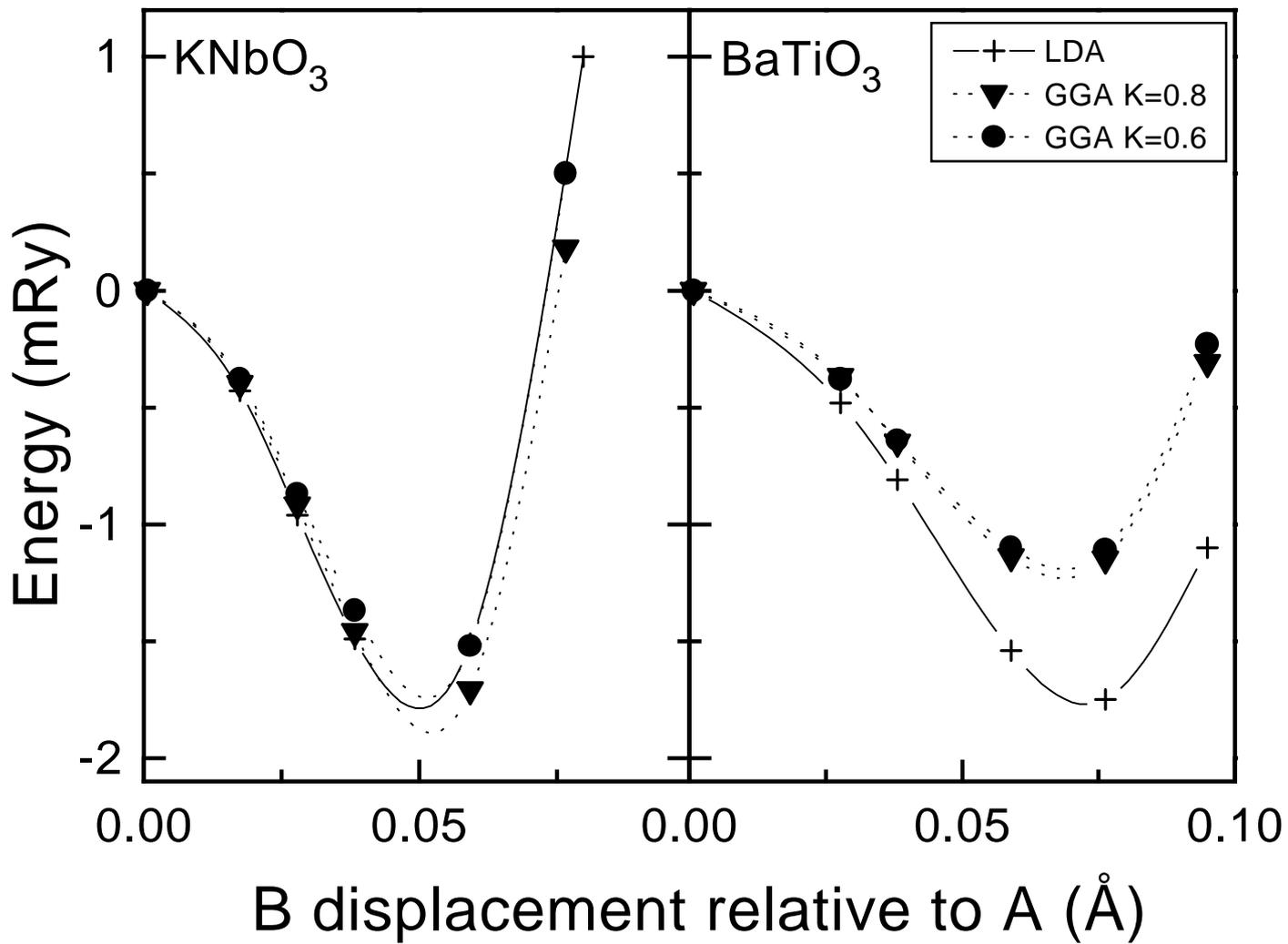
\end{document}